\documentclass{aastex6}
\usepackage{mathtools}

\newcommand{\seight}{S$_{8}$ }
\newcounter{chemequation}
\renewcommand*\thechemequation{[R\arabic{chemequation}]}
\newenvironment{chemequation}{%
\stepcounter{chemequation}%
\begin{equation}}%
{\tag*{\thechemequation}%
\end{equation}}

\begin{document}


\title{Sulfur Hazes in Giant Exoplanet Atmospheres: Impacts on Reflected Light Spectra}


\author{Peter Gao\altaffilmark{1,2,3}, Mark S. Marley, and Kevin Zahnle}
\affil{NASA Ames Research Center \\
Moffett Field, CA 94035, USA}

\author{Tyler D. Robinson\altaffilmark{4,5}}
\affil{Department of Astronomy and Astrophysics \\
University of California Santa Cruz \\
Santa Cruz, CA 95064, USA}

\and 

\author{Nikole K. Lewis}
\affil{Space Telescope Science Institute \\
Baltimore, MD 21218, USA}









\altaffiltext{1}{Division of Geological and Planetary Sciences, California Institute of Technology, Pasadena, CA 91125, USA}
\altaffiltext{2}{NASA Postdoctoral Program Fellow}
\altaffiltext{3}{pgao@caltech.edu}
\altaffiltext{4}{Sagan Fellow}
\altaffiltext{5}{NASA Astrobiology Institute's Virtual Planetary Laboratory}

\begin{abstract}

Recent work has shown that sulfur hazes may arise in the atmospheres of some giant exoplanets due to the photolysis of H$_{2}$S. We investigate the impact such a haze would have on an exoplanet's geometric albedo spectrum and how it may affect the direct imaging results of WFIRST, a planned NASA space telescope. For temperate (250 K $<$ T$_{\rm eq}$ $<$ 700 K) Jupiter--mass planets, photochemical destruction of H$_{2}$S results in the production of $\sim$1 ppmv of \seight between 100 and 0.1 mbar, which, if cool enough, will condense to form a haze. Nominal haze masses are found to drastically alter a planet's geometric albedo spectrum: whereas a clear atmosphere is dark at wavelengths between 0.5 and 1 $\mu$m due to molecular absorption, the addition of a sulfur haze boosts the albedo there to $\sim$0.7 due to scattering. Strong absorption by the haze shortward of 0.4 $\mu$m results in albedos $<$0.1, in contrast to the high albedos produced by Rayleigh scattering in a clear atmosphere. As a result, the color of the planet shifts from blue to orange. The existence of a sulfur haze masks the molecular signatures of methane and water, thereby complicating the characterization of atmospheric composition. Detection of such a haze by WFIRST is possible, though discriminating between a sulfur haze and any other highly reflective, high altitude scatterer will require observations shortward of 0.4 $\mu$m, which is currently beyond WFIRST's design. 

\end{abstract}




\section{Introduction} \label{sec:intro}

Observations of exoplanets have revealed a ubiquity of clouds and hazes that impede understanding of their composition. The presence of these clouds and hazes is typically shown by a flattening of spectral features in transmission spectra, resulting from the inability of stellar photons to reach depths in the atmosphere below the cloud and haze layers \citep{gibson2012,gibson2013,deming2013,jordan2013,mandell2013,sing2013,chen2014,schlawin2014,wilkins2014,fukui2014,mallonn2016}. Such flat transmission spectra have been seen across many exoplanets of different sizes, effective temperatures, and stellar irradiation levels \citep[e.g.][]{crossfield2013,kreidberg2014,knutson2014a,knutson2014b,sing2016}, suggesting that the processes governing cloud and haze formation in exoplanet atmospheres are as complex as they are ubiquitous. Clouds and hazes have also been inferred from optical phase curves, where higher than expected albedoes have been attributed to reflective clouds \citep[e.g.][]{demory2013}, and dust clouds have been invoked to fit spectra of directly imaged exoplanets \citep{barman2011,madhusudhan2011,marley2012,currie2014,skemer2014,chilcote2015,macintosh2015}. 

One key unknown is whether the aerosols blocking the stellar photons are part of a cloud, which condense from atmospheric gases and are typically supported in an atmosphere by turbulent mixing, or part of a haze, whose particles are usually produced by photochemistry and often found at high altitudes. \citet{morley2013} showed that, for the super Earth GJ 1214 b, photochemical hazes may be preferred as a solution to its flat transmission spectra, as they are formed high up in the atmosphere. By contrast, cloud particles must be lofted by turbulent mixing, and high metallicity may be required to ensure that enough material gets to the pressure levels probed by transmission spectroscopy.

The composition and vertical structure of atmospheric aerosols, whether clouds or hazes, is also of first order importance for understanding the reflected light spectra of a planet (see \citet{marley2013} for a review). Cloudless extrasolar giant planets, for example, are expected to be relatively bright at blue wavelengths but dark in the red optical and near infrared where molecular absorption dominates over Rayleigh scattering; cloudy planets, by scattering more photons before they can be absorbed, can be much brighter \citep{marley1999b,sudarsky2000}. Likewise for terrestrial planets \citet{morley2015} and \citet{charnay2015} showed that there are large differences in the reflected light spectra of super Earths depending on whether the planet is cloudy or hazy, and what kind of aerosols are present. Specifically, cooler planets with KCl and water clouds may be more reflective than clear planets in the near-infrared, while ZnS clouds, forming at similar temperatures, are more absorbing. In contrast, planets with complex hydrocarbon, ``soot'' hazes resulting from methane photolysis and polymerization could be very dark, whereas hazes composed of tholins--nitrogen rich organics thought to comprise Titan's hazes \citep{khare1984}--are dark at the blue end of the visible range and more reflective at the red end.

The reflectivity of clouds and hazes also impacts the amount of compositional information that can be retrieved from direct imaging, as reflected light is only able to sample the atmosphere above the cloud/haze deck. If the cloud/haze deck is at high altitudes, then the column optical depth of absorbing gases that direct imaging can sample is small, thereby reducing the magnitude of their spectral features in these planets' reflected light spectra \citep{marley1999b,sudarsky2000}. Meanwhile, if the cloud/haze is absorbing, then very few photons can sample the chemical composition of the atmosphere and escape. Such considerations are particularly relevant, as NASA is now studying future space based telescopes capable of characterizing extrasolar planets through high-contrast direct imaging \citep{krist2007,boccaletti2015,robinson2016}. 

An alternative set of compounds that are known to form hazes in planetary atmospheres are those derived from sulfur chemistry. In oxidizing atmospheres, SO$_{2}$, OCS, and H$_{2}$S from volcanic outgassing is transformed into sulfuric acid through photolysis and reactions with water. Condensation of sulfuric acid can then form clouds and hazes, such as the global cloud deck of Venus \citep{hansen1974} and the Junge layer in the upper stratosphere of Earth \citep{junge1963}. In reducing atmospheres, photolysis of H$_{2}$S can lead to the formation of elemental sulfur. For example, the existence of sulfur aerosols has been proposed for the Venus atmosphere, especially in regions depleted in oxidants \citep{toon1982,zhang2012}. Precipitation of elemental sulfur is the leading hypothesis for the preservation of mass independent fractionation of sulfur isotopes found in sediments on Earth older than 2.4 billion years \citep{pavlov2002}. \citet{hu2013} showed that terrestrial exoplanets with H$_{2}$--dominated atmospheres could be enveloped in optically thick sulfur hazes resulting from volcanic outgassing of H$_{2}$S. 

More recent work by \citet{zahnle2016} showed that rich sulfur photochemistry could potentially take place in the atmospheres of temperate giant exoplanets (250 K $<$ T$_{\rm eq}$ $<$ 700 K), generating sulfur allotropes that may condense to form hazes at lower temperatures. As these planets are the targets of current and planned direct imaging campaigns, it is essential that the optical characteristics of elemental sulfur hazes be known in order to inform these future observations. In this paper we investigate the geometric albedo spectra of giant planets hosting elemental sulfur hazes and their variations with haze properties, such as the location of the haze in the atmosphere and the haze optical depth. We also address the observability of a sulfur haze and its impact on inferences of atmospheric composition for the upcoming space--based direct imaging campaigns of WFIRST \citep{spergel2013}. In future publications, we will evaluate the impact of sulfur hazes on transmission spectroscopy and thermal emission, particularly from temperate, young directly imaged planets. 

In ${\S}$\ref{sec:sulfur}, we give an overview of the sulfur chemistry elaborated upon in \citet{zahnle2016}, with a focus towards the formation of sulfur hazes, as well as sulfur's optical properties. In ${\S}$\ref{sec:methods} we describe in brief the suite of models used in this study. In ${\S}$\ref{sec:results} we present our results showing how the geometric albedo varies with different haze properties and whether a clear planet can be distinguished from a hazy planet using proposed instruments onboard WFIRST. Finally, in ${\S}$\ref{sec:discussion} we discuss the impact of our assumptions, the implications of our results, and potential avenues of investigation for future missions and observation campaigns. 

\section{Sulfur in Giant Exoplanets}\label{sec:sulfur}

H$_{2}$S is the dominant reservoir of sulfur in giant exoplanet atmospheres under equilibrium chemistry for the temperatures and pressures considered here (T$_{\rm eq}$ $\sim$ 250--700 K, P $\sim$ 0.1--100 mbar). At temperatures below $\sim$200 K, NH$_{3}$ condenses and reacts with H$_{2}$S to form NH$_{4}$SH \citep{atreya1999,loeffler2015}, while above $\sim$1500 K (at P $<$ 1 bar) it dissociates into SH and S \citep{visscher2006}. The stability of H$_{2}$S is disrupted by disequilibrium processes, however, as upward transport of H$_{2}$S to the tropopause by turbulent mixing and advection results in its destruction by photolysis and reactions with atomic H released from photolysis of CH$_{4}$, NH$_{3}$, and H$_{2}$O \citep{zahnle2016}:

\begin{chemequation}
  \text{H}_2\text{S} + \text{H}  \longrightarrow \text{HS} + \text{H}_2
  \label{ch:chem1}
\end{chemequation}

\noindent The resulting HS radical quickly reacts to free sulfur and form S$_2$

\begin{chemequation}
  \text{HS} + \text{H}  \longrightarrow \text{S} + \text{H}_2
  \label{ch:chem2}
\end{chemequation}
\begin{chemequation}
  \text{HS} + \text{S}  \longrightarrow \text{H} + \text{S}_2
  \label{ch:chem3}
\end{chemequation}

\noindent This begins the polymerization process to form higher sulfur allotropes, eventually creating the stable allotrope, S$_{8}$. Transport of \seight into the deep atmosphere then results in its destruction via thermal decomposition

\begin{chemequation}
  \text{S}_8 + \text{M}  \longrightarrow 2\text{S}_4 + \text{M} 
  \label{ch:chem4}
\end{chemequation}

\noindent the products of which go on to decompose further before reforming H$_{2}$S, thus completing the cycle. In the event that the equilibrium \seight partial pressure is above the saturation vapor pressure $P_{\rm sat}$ of \seight somewhere in the atmosphere, given by \citep{zahnle2016}

\begin{equation}
P_{\rm sat} = \begin{cases}
		\exp{(20 - 11800/T)} \quad T < 413 \text{K} \\
		\exp{(9.6 - 7510/T)}  \quad T > 413 \text{K}
	       \end{cases}
\end{equation}

\noindent then an \seight haze may form. 

\begin{figure}[hbt!]
\centering
\includegraphics[width=0.6 \textwidth]{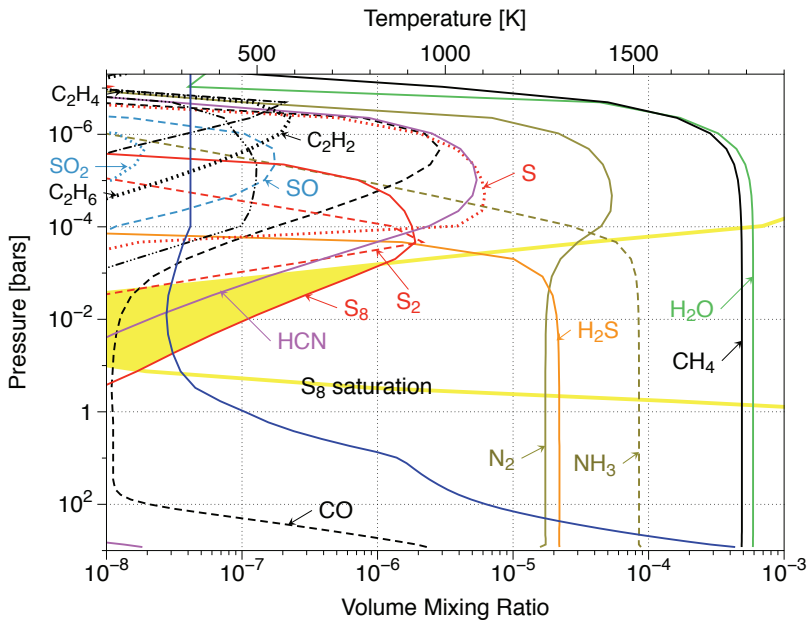}
\caption{The temperature profile (blue), \seight saturation vapor mixing ratio (yellow), and equilibrium mixing ratios of several important and/or sulfur--derived chemical species in a model giant exoplanet atmosphere subject to photochemistry and eddy diffusion. The shaded yellow region indicates where \seight is supersaturated. The temperature profile below 0.1 mbar is derived from a radiative--convective model (see ${\S}$\ref{sec:modatm}), with the temperature extending isothermally above that to include the necessary pressure levels important in photochemistry. }
\label{fig:zahnlechem}
\end{figure}

Figure \ref{fig:zahnlechem} shows the temperature profile (blue), saturation mixing ratio of \seight (saturation vapor pressure divided by total atmospheric pressure, in yellow), and the equilibrium mixing ratios of numerous chemical species in a temperate giant exoplanet atmosphere (see ${\S}$\ref{sec:modatm}) as a result of photochemistry and transport by eddy diffusion. The equilibrium mixing ratio of \seight peaks at $\sim$1 ppmv, and crosses the \seight saturation mixing ratio curve between 100 and 1 mbar (the yellow region). This allows for a rough estimate of the total mass of the haze, assuming that all \seight that is supersaturated condenses. For example, 1 ppmv of \seight at the 100 mbar level with $T$ $\sim$ 250 K results in a number density of \seight molecules of $\sim$3 $\times$ 10$^{12}$ cm$^{-3}$. Assuming a column height equaling one scale height ($\sim$20 km at 100 mbar for this atmosphere), then the column integrated number density of \seight is $\sim$6 $\times$ 10$^{18}$ cm$^{-2}$, which translates to a haze particle column number density of $\sim$3 $\times$ 10$^{11}$ cm$^{-2}$ assuming a particle size of 0.1 $\mu$m and a mass density of 2 g cm$^{-3}$. 

The ultimate haze mass will depend on the microphysics of haze formation (see ${\S}$\ref{sec:discussion}) and the degree to which \seight is supersaturated, which in turn depends on the equilibrium \seight mixing ratio and the \seight saturation vapor mixing ratio. \citet{zahnle2016} showed that, for a wide range of stellar UV fluxes and eddy diffusivities, the peak equilibrium \seight mixing ratio remained close to 1 ppmv to within a factor of 2, though its vertical profile became more extended/compressed when the eddy diffusivity increased/decreased, respectively. The \seight mixing ratio is largely independent of stellar UV fluxes because the fluxes experienced by temperate giant exoplanets are such that the photochemistry is limited by H$_{2}$S upwelling, rather than the supply of UV photons. The \seight mixing ratio is also independent of the eddy diffusivity because, in the event that \seight does not condense, the upward mixing of H$_{2}$S is balanced by the downward mixing of S$_{8}$, and thus changing the rate of mixing should not change the size of the reservoirs from which material is exchanged by mixing. Note however that as precipitation of sulfur becomes important, the mass of haze will begin to reflect the UV flux and the vertical mixing; we will revisit this topic in future work. Furthermore, as Figure \ref{fig:zahnlechem} shows, changing the temperature does not greatly alter the \seight mixing ratio because to first approximation all the sulfur is in the form of \seight molecules. The reason \seight is favored is that all other S-containing molecules are easily photolyzed by abundant 200-300 nm UV photons. \seight is also subject to photolysis, but it is a ring molecule, which means that after photolysis it becomes a linear \seight molecule that is rather likely to grab its tail and re-emerge as a ring. In other words, though varying the temperature changes the rates of reactions, all reactions eventually lead to the transformation of H$_{2}$S to \seight above the tropopause.

\begin{figure}[hbt!]
\centering
\includegraphics[width=0.6 \textwidth]{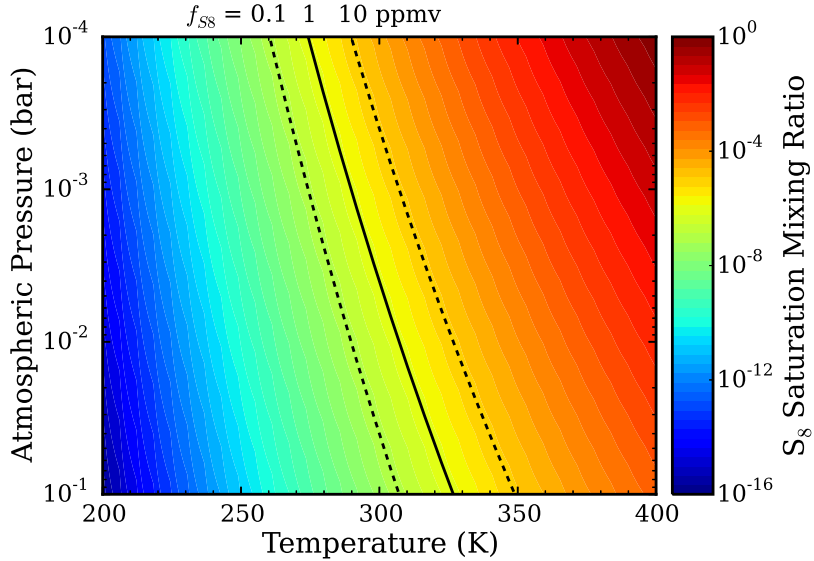}
\caption{\seight saturation vapor mixing ratio as a function of temperature and background atmospheric pressure. The solid line indicates where the \seight saturation vapor mixing ratio equals 1 ppmv, while the dotted lines to the left and right indicate 0.1 and 10 ppmv, respectively.}
\label{fig:s8svp}
\end{figure}

Given the stability of the equilibrium \seight mixing ratio, the \seight supersaturation will largely depend on the saturation vapor mixing ratio, which is a function of temperature. Figure \ref{fig:s8svp} shows the \seight saturation vapor mixing ratio as a function of temperature and pressure level in the atmosphere, where the range in pressure level denotes where \seight tends to be abundant \citep{zahnle2016}. The solid line indicates a saturation vapor mixing ratio of 1 ppmv, while the dashed lines to the left and right side of it indicate 0.1 and 10 ppmv, respectively. Thus, if an exoplanet atmosphere contains 1 ppmv of \seight, then condensation can occur (\seight is supersaturated) for temperatures and atmospheric pressure levels to the left of the 1 ppmv line, while to the right the abundance of \seight is too low to condense. As previously indicated, however, NH$_{4}$SH clouds are expected to form at temperatures $<$200 K (when NH$_{3}$ may condense), which removes H$_{2}$S from the atmosphere and arrests the photochemical production of S$_{8}$. Therefore, we can conclude that sulfur hazes may arise on temperate giant exoplanets with stratospheric temperatures $<$325K but warmer than Jupiter, assuming that their metallicity is solar. Increasing the metallicity leads to an increase in the high temperature limit for sulfur haze formation (and vice versa for lower metallicity) due to the increased sulfur mixing ratio in the atmosphere, i.e. if the equilibrium \seight mixing ratio is 10 ppm rather than 1 ppm, then sulfur hazes can form up to $\sim$350 K. The situation at the low temperature limit is more uncertain, however, due to the unknowns in the formation of NH$_{4}$SH clouds.

\begin{figure}[hbt!]
\centering
\includegraphics[width=0.6 \textwidth]{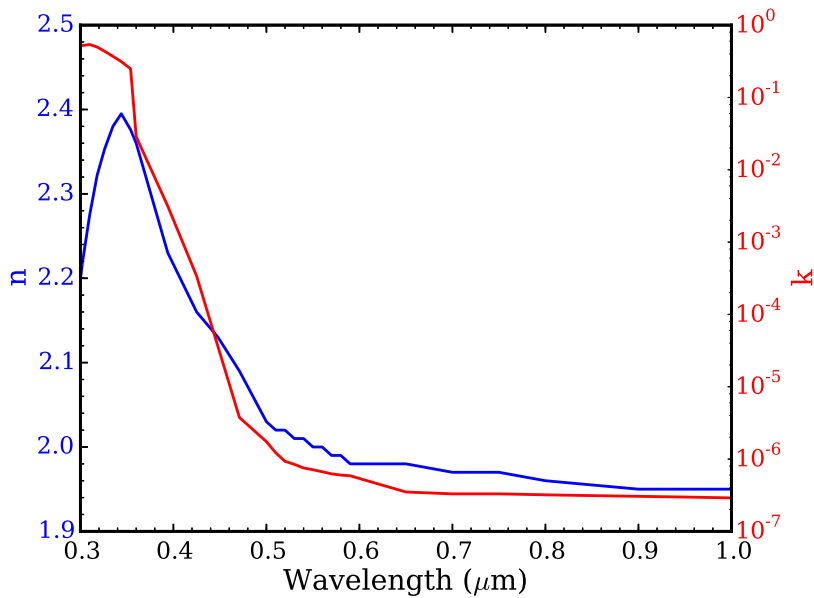}
\caption{Real (blue) and imaginary (red) components of S$_{8}$'s complex refractive index \citep{fuller1998}}
\label{fig:s8nk}
\end{figure}

The effect of a sulfur haze on a planet's geometric albedo depends heavily on sulfur's optical properties, stemming primarily from its complex refractive index. Figure \ref{fig:s8nk} shows the real ($n$) and imaginary ($k$) indices of refraction for orthorhombic crystals of sulfur, a form of solid sulfur composed mainly of S$_{8}$. Orthorhombic sulfur is the most thermodynamically stable form of solid \seight for the temperatures relevant here and is therefore the most likely composition of sulfur hazes on giant exoplanets, with other forms converting to orthorhombic sulfur on finite time scales that tend to decrease with increasing temperatures \citep{fuller1998}. Above 368 K, an alternative form, monoclinic sulfur, becomes thermodynamically preferred \citep{meyer1976}. A key feature of orthorhombic sulfur's complex refractive index is the increase in $k$ at shorter wavelengths caused by vibrational modes in the \seight molecules, which tend to become more populated at higher temperatures, thereby shifting the increase in $k$ to longer wavelengths \citep{meyer1972}. Increased absorption at shorter wavelengths will lead to sulfur haze--enveloped exoplanets being yellowish to reddish in color. 

\section{Methods}\label{sec:methods}

\subsection{Model Atmosphere}\label{sec:modatm}

We evaluate the effect a sulfur haze would have on a planet's geometric albedo spectrum by introducing a sulfur haze into a 1--dimensional model background atmosphere, which we initially assume to be devoid of clouds or hazes. The pressure--temperature profile of the model atmosphere is set by asserting radiative--convective equilibrium, and its molecular composition is determined by assuming thermochemical equilibrium. To illustrate the effect of the haze we use the planetary parameters of Gamma Cephei Ab, a warm giant exoplanet discovered via radial velocity surveys \citep{campbell1988,hatzes2003} with a minimum mass of 1.85 Jupiter masses, and which orbits its host star, a K1IVe subgiant with an effective temperature of 4800 K and a stellar radius of 4.9 solar radii, at a semimajor axis of 2.05 AU \citep{torres2007,endl2011}. We choose Gamma Cephei Ab due to its moderate stratospheric temperatures ($\sim$260 K), which could be conducive to sulfur haze formation, as well as its plausibility as a target for WFIRST \citep{lupu2016}. 

The PT profile of our atmosphere is generated using the iterative radiative--convective model developed in \citet{mckay1989}, and extended by \citet{marley1996,marley1999,marley2002,burrows1997,fortney2005,fortney2008,saumon2008}. Given an internal heat flux and incident flux from the host star, the PT profile is adjusted until (1) the net flux between the plane--parallel atmospheric layers is zero and (2) the profile adheres to convective stability. The radiative transfer is treated via the two--stream source function method described in \citet{toon1989}, with opacities of molecular species provided by \citet{freedman2008} with updates from \citet{saumon2012} and combined using the correlated--k method \citep{goody1989}. Figure \ref{fig:zahnlechem} shows the model atmosphere pressure--temperature profile in blue, where only the regions below 0.1 mbar are in radiative--convective equilibrium, with the radiative/convective boundary at $\sim$0.1 bar. Extension of the PT profile upwards was necessary to investigate the abundance of \seight produced in this model atmosphere due to photochemistry. The assumption that the PT profile extends upwards isothermally should not affect our results significantly, since (1) we do not consider altitudes above the 0.1 mbar level in our geometric albedo calculations, (2) \seight production is largely independent of temperature (${\S}$\ref{sec:sulfur}), and (3) most of the haze formation occurs at altitudes below the 0.1 mbar level. 

We assume solar metallicity for the model atmosphere \citep{lodders2003}, and calculate its molecular composition by minimizing the Gibbs free energy to ensure thermochemical equilibrium. The major chemical species that are abundant and/or optically active include H$_{2}$, H, VO, TiO, CO$_{2}$, He, H$_{2}$O, CH$_{4}$, CO, NH$_{3}$, N$_{2}$, PH$_{3}$, H$_{2}$S, Fe, Na, and K. In the event that a species becomes supersaturated, it is assumed to be depleted via condensation above the crossing point between its partial pressure and its saturation vapor pressure \citep{lodders1999}. The thermochemical equilibrium composition is used to calculate the planet's geometric albedo, and serve as initial conditions for the time--stepping photochemical model, the results of which are shown in Figure \ref{fig:zahnlechem}. 

\subsection{Cloud Model and Haze Treatment}\label{sec:cloudhaze}


The PT profile of our model atmosphere is such that KCl and ZnS can potentially condense and form clouds \citep{morley2012}. To include clouds in our model we use the approach of \citet{ackerman2001}, where the cloud mass and mean particle size is determined by balancing vertical mixing due to eddy diffusion and sedimentation of cloud particles

\begin{equation}
K_{zz}\frac{\partial q_{\rm t}}{\partial z} + f_{\rm sed} w q_{\rm c} = 0
\end{equation}

\noindent where $q_t$ is the total mixing ratio of the condensing species, $q_c$ is the mixing ratio of the condensed form of the condensing species only, $w$ is the convective velocity, $f_{sed}$ is a tunable parameter and a measure of the sedimentation efficiency of the cloud particles that we set to 3, appropriate for Jupiter--like words \citep{ackerman2001,saumon2008,morley2012}, and $K_{zz}$ is the eddy diffusion coefficient, calculated by assuming that the convective mass transport is equivalent to convective heat transport. A minimum value $K_{zz}^{min}$ = 10$^{5}$  cm$^{2}$ s$^{-1}$ is set in regions that are convectively stable, as eddy diffusion does not only represent convective turbulence. $w$ is then calculated via mixing length theory: $w$ = $K_{zz}/L$, where $L$ is the mixing length given by 

\begin{equation}
L = H\max (0.1, \Gamma/\Gamma_{ad} )
\end{equation}

\noindent where $H$ is the atmospheric scale height, $\Gamma$ and $\Gamma_{ad}$ are the local and dry adibatic lapse rates, and 0.1 is the minimum scaling for $L$ when the atmosphere is convectively stable. 

For both the clouds and sulfur haze, we assume that the particle size distribution $dn/dr$ is lognormal, given by

\begin{equation}
\frac{dn}{dr} = \frac{N}{r\sqrt{2\pi} \ln{\sigma}} \exp \left [ - \frac{\ln^2{\left (r/r_g \right )}}{2 \ln^2{\sigma}} \right ]
\end{equation}

\noindent where $N$ is the total number density of particles, $r$ is the particle radius, $r_g$ is the mean particle radius, and $\sigma$ is a measure of the width of the distribution and is fixed to 2 in the cloud model. This choice of $\sigma$ is motivated in \citet{ackerman2001} as it is broad enough to account for both freshly condensed particles and those which have grown by coagulation. We plan to explore sulfur particle growth and sedimentation more fully in a future study.

We position our nominal sulfur haze layer at the 10 mbar level in the model atmosphere, consistent with the results of \citet{zahnle2016} and our Figure \ref{fig:zahnlechem}; the column number density of haze particles is set to 10$^{11}$ cm$^{-2}$, in agreement with our calculations of the haze mass in ${\S}$\ref{sec:sulfur}; and the mean particle radius is set to 0.1 $\mu$m, similar to mode 1 particles in the clouds of Venus, which may be composed of elemental sulfur \citep{knollenberg1980}. We will test the sensitivity of geometric albedo spectra to these parameters. 

The optical properties of the clouds and the sulfur haze, such as their optical depth, single scattering albedo, and asymmetry parameter (a measure of their degree of forward scattering) are calculated by the cloud model using Mie theory assuming homogeneous spheres. As with the sulfur haze, the optical properties of the clouds are determined by their complex indices of refraction, which for KCl and ZnS are provided by \citet{querry1987}. 

In a self--consistent atmosphere, the formation of clouds and hazes would perturb the PT profile, which will in turn lead to different molecular abundances. However, for this exercise we do not ensure this self-consistency. Instead, the PT profile is fixed to that of a clear atmosphere (though the molecular abundances do reflect condensation), with condensate clouds and the prescribed sulfur haze layer added to it afterwards. We discuss the consequences of this assumption in ${\S}$\ref{sec:discussion}.






\subsection{Geometric Albedo Model}\label{sec:geomalb}

The geometric albedo of a planet is defined as the ratio of the reflected flux of that planet at full phase to the reflected flux from a perfect Lambert disk with the same radius as the planet located at the same distance from its host star. Our geometric albedo model is based on that developed for Titan by \citet{mckay1989} and extended to solar and extrasolar giant planets by \citet{marley1999,marley1999b} and \citet{cahoy2010}. The reflecting hemisphere of our cloudy/hazy planet is split into 100 individual 1--dimensional atmospheric columns with plane--parallel layers, and radiative transfer calculations for each column is performed separately following \citet{toon1989}, relating incident fluxes to reflected fluxes. At full phase, each patch on the planet has the same observer and solar zenith angle, and these vary across the disk due to the curvature of the reflecting hemisphere. The geometric albedo spectrum is then calculated by integrating over the reflected flux from each column weighted by viewing geometry at full phase \citep{cahoy2010}.

Opacity sources considered when calculating the geometric albedo spectrum include molecular absorption as previously mentioned (including collision--induced absorption of H$_{2}$--H, H$_{2}$--H$_{2}$, H$_{2}$--He, and H$_{2}$--CH$_{4}$), as well as Rayleigh scattering \citep{cahoy2010} and Raman scattering \citep{pollack1986}. The optical depths, single scattering albedos, and asymmetry parameters calculated by the cloud model are fed as input parameters into a double Henyey-Greenstein phase function to account for the opacity and anisotropic scattering of the condensate clouds and the sulfur haze. 

\subsection{WFIRST Noise Model}\label{sec:wfirstmod}

WFIRST (Wide Field Infrared Survey Telescope) is a planned NASA mission to be launched in the mid--2020s. It is a space--based, coronagraph--equipped observational platform with a 2.4 m diameter primary aperture, an operational wavelength range from $\sim$0.4 to 1 $\mu$m, and a spectral resolution of 70 \citep{spergel2013}. \citet{robinson2016} showed that, given a coronagraph capable of achieving a planet--star contrast ratio of 10$^{-9}$ and expected levels of read noise, dark current, leaked stellar light, and zodiacal light from the Solar System and the exoplanetary system (noise terms that can be comparable to, or even dominate the planetary signal itself), a giant exoplanet located at 2 AU from a sun--like star is readily detectable and characterizable with integration times of several tens of hours, provided that the exoplanetary system is located within about 10 pc and that the planet--star angular separation is within the coronagraph inner and outer working angles. \citet{lupu2016} expanded beyond these constraints and gave a preliminary target list for WFIRST (see their Figure 1), which showed several objects ($\sim$1/3) within the temperature range given in ${\S}$\ref{sec:sulfur}. Therefore, we can expect a significant fraction of WFIRST targets to exhibit atmospheric conditions favorable to sulfur haze formation. 

We investigate the observability of a sulfur haze and its impact on the detectability of various atmospheric chemical species using an updated version of the noise model described in 
\citet{robinson2016}, which includes specific coronagraph designs that have been proposed for WFIRST, such as the Shaped--Pupil Coronagraph \citep[SPC; capable of both broadband imaging and spectroscopy;][]{kasdin2004} and the Hybrid Lyot Coronagraph \citep[HLC; imaging mode only;][]{trauger2016}. The baseline telescope and instrument parameter values corresponding to these coronagraphs are different from that of \citet{robinson2016} (see their Table 3). We show these updated values in Table \ref{tab:wfirstnewparams}. 

\begin{deluxetable}{llccc}
\tablecaption{Changes to Telescope and Instrument Parameter Values\label{tab:wfirstnewparams}}
\tablehead{
 & & \colhead{Value from} & \colhead{Value of} & \colhead{Value of} \\ 
\colhead{Parameter} & \colhead{Description} & \colhead{\citet{robinson2016}} & \colhead{Current Work (HLC)} & \colhead{Current Work (SPC)} 
}
\startdata
$D$ & telescope diameter & 2 m & 2.4 m  & 2.4 m\\
$R_{e-}$ & read noise counts per pixel & 0.1 & 0.2 & 0.2\\
$\mathcal{T}$ & telescope and instrument throughput & 0.05 & 0.074 & 0.037 \\
$\theta_{IWA}$ ($\lambda$/D) & coronagraph inner working angle & 2 & 2.8 & 2.7 \\
\enddata
\tablecomments{Only parameters updated for this work are shown.}
\end{deluxetable}

To fully characterize the impact of sulfur hazes on the geometric albedo spectrum, we position our exoplanetary system, composed of the K1IVe subgiant host star and the warm giant exoplanet at 8 pc distance so that the full wavelength range under consideration can be observed without going beyond the inner or outer working angles (the real Gamma Cephei Ab is located 13.79 pc away \citep{hatzes2003}, such that wavelengths longer than $\sim$0.6 $\mu$m would not be observable due to the planet--star on--sky separation for those wavelengths being within WFIRST's inner working angle). In addition, we assume that (1) the exozodi brightness is the same as zodiacal light in our own Solar System and (2) our model planet has the same radius as Jupiter. As the observation would likely be made while the exoplanet is at quadrature rather than at full phase, we multiply the contrast ratio calculated from the (full phase) geometric albedo by a correction factor 1/$\pi$, roughly simulating the drop in brightness at quadrature versus full phase. However, this does not take into account the changes in geometric albedo due to non--uniform scattering phase functions of the clouds and hazes in the atmosphere. It should also be noted that the real Gamma Cephei A is part of a binary system, which would reduce WFIRST's ability to reach its designed planet--star raw contrasts. For simplicity, we assume that our model host star does not have a stellar companion. 

\section{Results} \label{sec:results}

\subsection{Geometric Albedo Spectra}\label{sec:gas}

\begin{figure}[hbt!]
\centering
\includegraphics[width=0.6 \textwidth]{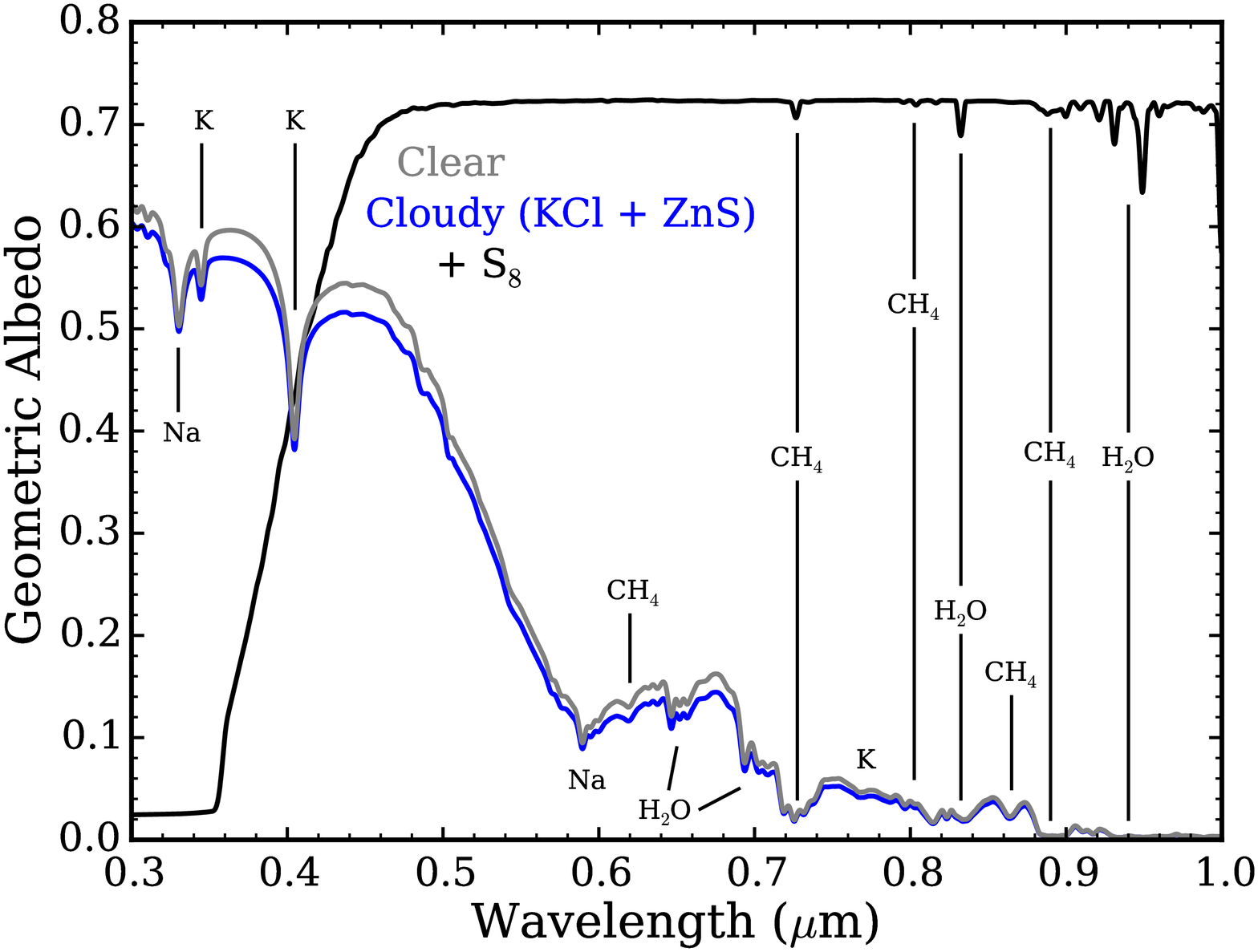}
\caption{Geometric albedo spectra for a clear (gray), cloudy (blue), and hazy (black) giant exoplanet atmosphere. The cloudy case includes KCl and ZnS clouds. The hazy case includes all aforementioned clouds, and the nominal sulfur haze layer located at 10 mbar with a column particle number density of 10$^{11}$ cm$^{-2}$ and a mean particle size of 0.1 $\mu$m. The major absorption features are labeled \citep{cahoy2010,morley2015}.}
\label{fig:nominalresults}
\end{figure}


Figure \ref{fig:nominalresults} shows the geometric albedo spectra of a clear Gamma Cephei Ab atmosphere (gray), an atmosphere with KCl and ZnS clouds included (blue), and our nominal hazy model including both the aforementioned clouds and a sulfur haze (black). Absorption features due to H$_{2}$O and CH$_{4}$ can be clearly discerned at longer wavelengths in the clear case, which makes the planet especially dark there. At shorter wavelengths Rayleigh scattering increases the geometric albedo, with absorption dominated by alkali metals such as potassium and sodium \citep{cahoy2010,morley2015}. Adding condensation clouds darkens the albedo by $\sim$5\% due to ZnS absorption, even though the metallicity is not high enough to produce optically thick KCl and ZnS clouds.

In contrast, adding a sulfur haze layer drastically alters the geometric albedo spectrum. The CH$_{4}$ and H$_{2}$O absorption bands are now mostly absent save for some small dips longward of 0.7 $\mu$m, where the planet is much brighter due to the purely scattering sulfur haze there, while at wavelengths $<$0.45 $\mu$m the albedo drops precipitously due to increased sulfur absorption, and all traces of alkali metal absorption are gone due to their depletion in the upper atmosphere from KCl and ZnS cloud formation deeper in the atmosphere. These features are strongly reminiscent of the geometric albedo spectrum of Venus, where sulfuric acid provides the high albedo at long wavelengths, and an unknown UV absorber decreases the albedo shortward of $\sim$0.45 $\mu$m, though not as severely as what is seen here \citep[the albedo shortward of 0.45 $\mu$m and longward of the SO$_{2}$ absorption band at 0.3 $\mu$m is $\sim$0.5;][]{pollack1980}. While \seight has been proposed as a candidate for the unknown absorber \citep{hapke1975,young1977}, atmospheric models that included \seight as one of the cloud particle modes were unable to fit the observations \citep{pollack1979}. However, these studies do not rule out more complex scenarios that incorporate S$_{8}$. For example, the ``gumdrop model'' supposes that the Venus cloud particles may be layered spheres with cores of sulfuric acid surrounded by a thin shell of sulfur arising from collisions with small sulfur aerosols or direct condensation of sulfur onto the cores \citep{young1983,markiewicz2014,petrova2015}. Furthermore, reactions between sulfur and sulfuric acid can yield compounds of intermediate redox states that absorb UV, such as thiosulfates in the oxidizing Venus atmosphere, and sulfanes in reducing atmospheres. 

The shape of the geometric albedo spectrum, as produced by a sulfur haze, also contrasts with those produced by hazes composed of other materials. Hydrocarbon hazes such as soots, for example, darken the planet across the entire wavelengths range studied here, while tholins are more similar to sulfur in that they also brighten a planet at long wavelengths while darkening it at short wavelengths, though the transition from high to low albedo for a tholin haze is considerably more gradual \citep{morley2015}. Neither of these types of hazes are relevant for the kind of planets considered here, however, as hydrocarbon hazes are likely in low abundance at the relevant pressure levels due to reactions between haze precursors with OH \citep{zahnle2016}. 

\begin{figure}[hbt!]
\centering
\includegraphics[width=0.6 \textwidth]{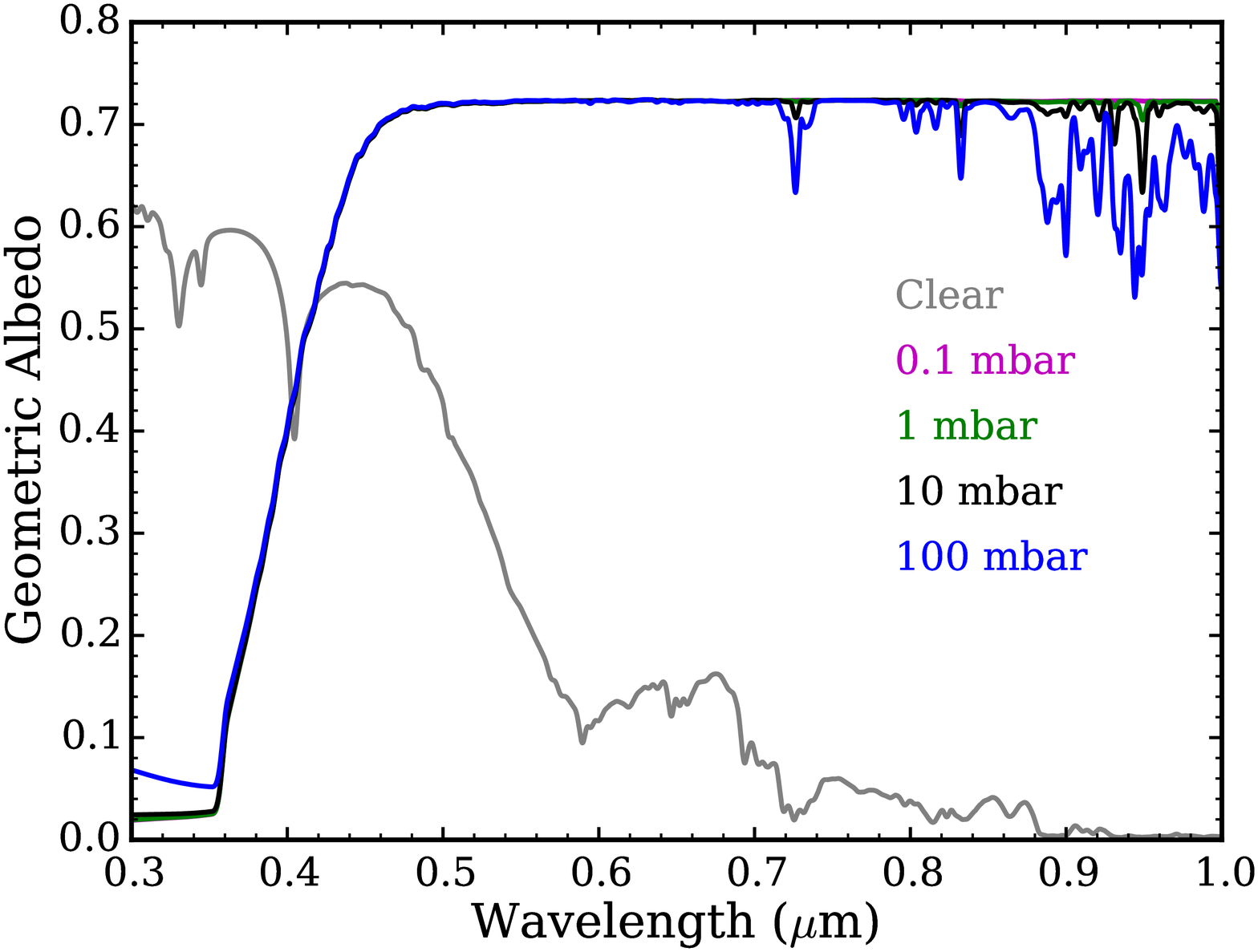}
\caption{Geometric albedo of a giant exoplanet with a sulfur haze located at 0.1 (magenta), 1 (green), 10 (black, nominal value), and 100 mbar (blue). The geometric albedo of a clear atmosphere (gray) is shown for comparison.}
\label{fig:prelevresults}
\end{figure}

Figure \ref{fig:prelevresults} shows the variations in the geometric albedo spectrum as the sulfur haze layer is placed at different pressure levels in the atmosphere, within the range of pressure levels where \seight is abundant. The location of the sulfur haze layer can be variable since it depends on where the \seight mixing ratio curve intercepts the \seight saturation vapor mixing ratio curve, which in turn is a function of the atmospheric temperature structure and intensity of vertical mixing. There is very little difference between the different cases, with the only variations due to increased absorption by CH$_{4}$ and H$_{2}$O as the haze is lowered in the atmosphere, thereby increasing the optical depth of these gases above the haze. 

\begin{figure}[hbt!]
\centering
\includegraphics[width=0.6 \textwidth]{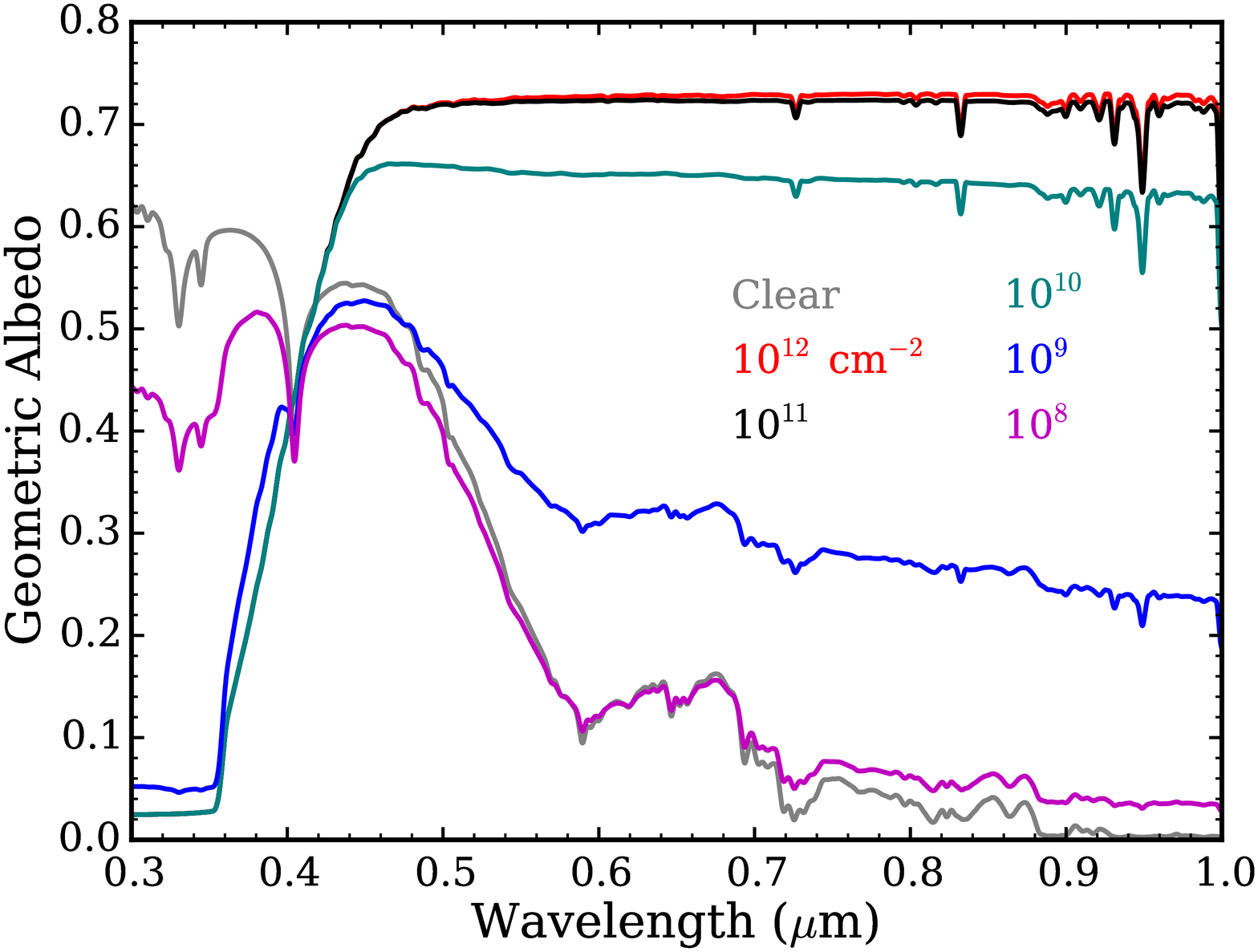}
\caption{Geometric albedo of a giant exoplanet with a sulfur haze with column particle number densities of 10$^{12}$ (red), 10$^{11}$ (black, nominal value), 10$^{10}$ (teal), 10$^{9}$ (blue), and 10$^{8}$ cm$^{-2}$ (magenta). The geometric albedo of a clear atmosphere (gray) is shown for comparison.}
\label{fig:numdenresults}
\end{figure}

Figure \ref{fig:numdenresults} shows the changes in the geometric albedo spectrum as the column number density of sulfur haze particles is varied. The column number density is strongly related to the degree of supersaturation of \seight vapor and is controlled largely by the microphysics of sulfur haze formation. Increasing the column number density from our nominal case does not change the geometric albedo spectrum to any large degree, indicating that the effect of the sulfur haze has already ``saturated'' for our nominal haze abundance. This is not surprising since the optical depth of our nominal case is already $>$1. Decreasing the column number density past an optical depth of 1 reduces the effect of the haze on the geometric albedo. In particular, the brightness of the planet is reduced significantly longward of 0.45 $\mu$m until it begins to match the clear case. By contrast, the geometric albedo shortward of 0.45 $\mu$m remains largely unchanged even at very low sulfur haze optical depths, only approaching the clear case for optical depths 1000 times less than that of the nominal case. 

\begin{figure}[hbt!]
\centering
\includegraphics[width=0.6 \textwidth]{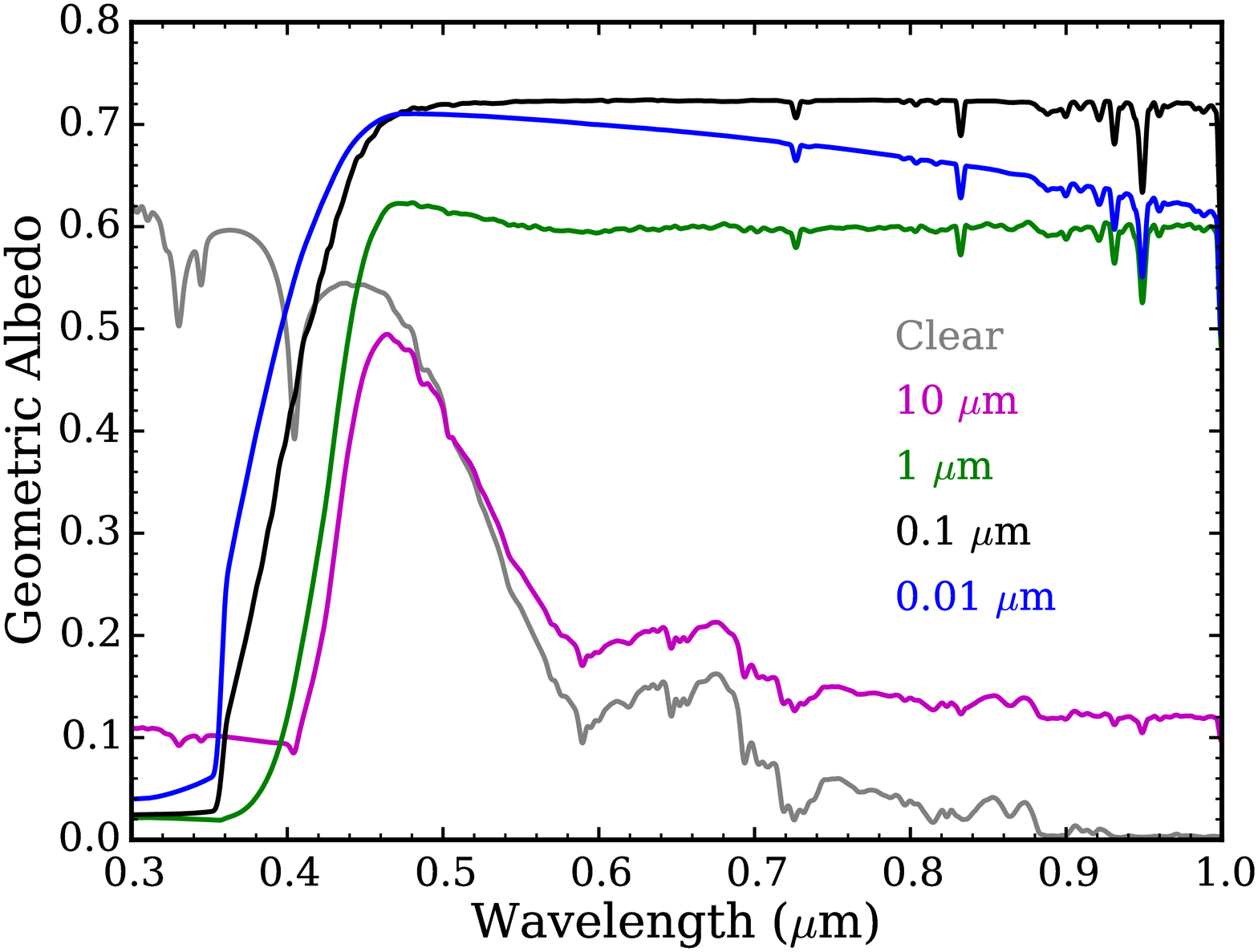}
\caption{Geometric albedo of a giant exoplanet with a sulfur haze with a mean particle radius of 0.01 (blue), 0.1 (black, nominal value), 1 (green), and 10 $\mu$m (magenta). The total haze mass is kept constant for all cases. The geometric albedo of a clear atmosphere (gray) is shown for comparison.}
\label{fig:partradresults}
\end{figure}

Figure \ref{fig:partradresults} shows the changes in the geometric albedo spectrum as the mean sulfur haze particle size is varied, while keeping the total haze mass the same (i.e. increasing particle size leads to a lower column particle number density). Like the column number density, the particle size depends on the microphysics of sulfur haze formation and growth by condensation of \seight vapor. Varying the particle size is seen to have different effects depending on whether the particles are mostly scattering (longward of 0.45 $\mu$m) or absorbing (shortward of 0.45 $\mu$m). In the scattering region, decreasing the particle size such that it becomes much smaller than the wavelength results in a Rayleigh slope developing at longer wavelengths, darkening the planet slightly. Increasing the particle size past the considered wavelength range leads to a decrease in the geometric albedo due to the decrease in haze optical depth. This results from the consolidation of haze mass in larger particles, since optical depth is proportional to the square of the particle radius, while particle mass is proportional to the cube of the particle radius, and the scattering efficiency is largely independent of the particle radius for radii much greater than the wavelength \citep{hulst1957}. The geometric albedo is largely unchanged in the absorbing part of the spectrum. This can be explained by the roughly linear relationship between particle size and the absorption efficiency \citep{hulst1957}; in other words, the decrease in haze optical depth due to consolidating mass in larger particles is balanced by an increase in absorption by those larger particles. In addition, the wavelength at which the geometric albedo drops abruptly moves to longer wavelengths with increasing particle size. This is because larger particles are more absorbing, and smaller particles are more Rayleigh scattering, which increases the albedo at shorter wavelengths. 
 
\subsection{Impact on WFIRST Observations}\label{sec:wfirst}

\begin{figure}[hbt!]
\centering
\includegraphics[width=0.6 \textwidth]{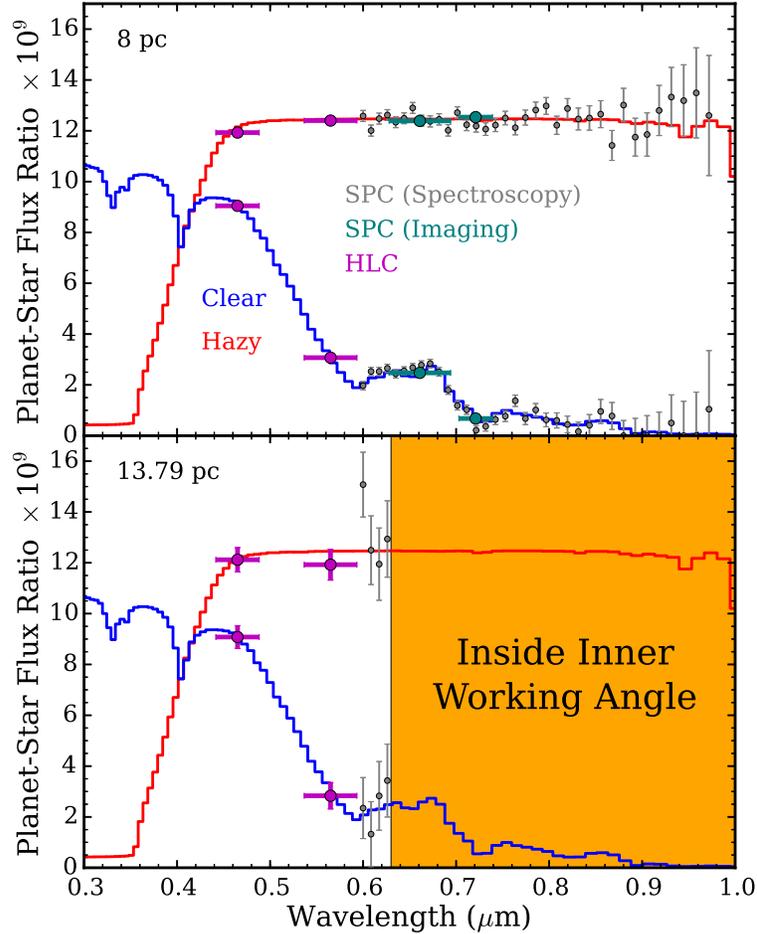}
\caption{Planet--star flux ratio $\times$ 10$^{9}$ for a clear (blue) and hazy (red) giant exoplanet with the same radius as Jupiter orbiting Gamma Cephei A -- a K1IVe subgiant with a radius of 4.9 $\times$ solar radii and an effective temperature of 4800 K -- at 2.05 AU, placed 8 pc away (top) and at its actual distance, 13.79 pc (bottom). Synthetic data from the shaped pupil coronagraph (SPC) in imaging (teal) and spectroscopy (gray) mode and the hybrid Lyot coronagraph (HLC; magenta), possible instruments onboard WFIRST, are overplotted. Integration time for each exposure is set to 20 hours, with the SPC in spectroscopy mode needing 3 exposure to cover the full wavelength range presented in the figure. In the case where the system is located at its actual distance, wavelengths longward of $\sim$0.63 $\mu$m are inside the inner working angle, and thus cannot be observed by WFIRST in its current (planned) configuration.}
\label{fig:wfirstresults}
\end{figure}

We can assess whether the drastic changes to a giant exoplanet's geometric albedo caused by a sulfur haze layer is detectable using the updated WFIRST noise model. Figure \ref{fig:wfirstresults} shows the low resolution (R = 70) planet--star flux ratio for the clear atmosphere case (blue) and the nominal hazy case (red). Superimposed on the spectra are synthetic observations taken by the shaped pupil coronagraph (SPC) in imaging (teal) and spectroscopy (gray) mode, and the hybrid Lyot coronagraph (HLC; magenta), each with a 20 hour integration time per observation. Note that the shaped pupil coronagraph in spectroscopy mode requires three exposures to capture its full wavelength range, so the full observation time to obtain SPC spectra is 60 hours. The larger error bars near 1 $\mu$m are due to the dropping quantum efficiencies of the detector \citep[$\sim$90\% at 0.6 $\mu$m vs. $\sim$10\% at 1 $\mu$m;][]{traub2016,robinson2016}. The top figure shows our idealized case, where the star system is placed at 8 pc to ensure that all wavelengths under consideration can be observed. The bottom figure shows the case where the star system is placed at its actual distance of 13.79 pc \citep{hatzes2003}.

There is a significant difference between the clear and hazy cases. For the wavelength range WFIRST is able to observe in the 8 pc case, the hazy planet is brighter by a factor of $\sim$6, and thus easier to detect. On the other hand, the minor CH$_{4}$ and H$_{2}$O absorption features in the hazy spectra are not detectable, thus rendering any effort to retrieve their abundances hopeless \citep{lupu2016}. This is even more so in the case where the star system is placed as its actual distance, where the molecular features are not even accessible since they are at wavelengths within the inner working angle of the coronagraph. This does not affect the HLC channels at shorter wavelengths, however, which in both cases show an obvious Rayleigh slope for the clear planet. In contrast, the hazy planet has a geometric albedo spectrum that is flat for all observable wavelengths, making clear and hazy atmospheres easily distinguishable even for the actual Gamma Cephei Ab. 

Nevertheless, a flat spectrum does not in itself confirm the existence of a sulfur haze, as its defining feature is its strong absorption shortward of 0.45 $\mu$m, which WFIRST cannot observe. However, while \citet{morley2015,charnay2015} showed that high water ice clouds in a colder atmosphere and KCl and ZnS clouds in a warmer atmosphere could also produce flat geometric albedo spectra, the metallicities necessary ($>$50 $\times$ solar) are far higher than what has been observed in giant planets. In fact, \citet{macdonald2017} showed that the geometric albedo spectrum of a giant exoplanet with water clouds is replete with molecular features. Therefore, the existence of sulfur hazes offers a likely explanation for any future observations of temperate giant exoplanets with flat geometric albedo spectra longward of 0.45 $\mu$m.

\section{Discussion} \label{sec:discussion}

We have demonstrated that sulfur photochemistry in temperate giant exoplanets with solar metallicity inexorably generates $\sim$1 ppmv of \seight vapor for large swaths of planetary parameter space. Therefore, a sulfur haze could form if the stratospheric temperature is low enough such that \seight becomes supersaturated. The effect of such a sulfur haze on the planet's geometric albedo is significant: pure scattering sulfur particles boost the albedo longward of 0.45 $\mu$m to $\sim$0.7, while strong absorption shortward of 0.45 $\mu$m lowers the albedo to near zero. This is the opposite trend to a clear atmosphere, where deep molecular absorption darkens a planet at long wavelengths while Rayleigh scattering brightens the planet at short wavelengths. Therefore, the existence of a sulfur haze would drastically change the color of the planet, shifting it from blue to orange. The haze also vastly reduces the detectability of atmospheric chemical species such as methane and water, as most of their absorbing column density lie below the sulfur haze. As a result, WFIRST direct imaging targets that exhibit sulfur hazes are unlikely to provide useful constraints on their atmospheric composition.  

Our results depend heavily on the properties of the haze, such as the mean particle size and column particle number density; these are determined by microphysical processes, such as the nucleation of aerosols, growth by condensation and coagulation, loss through evaporation and collisional breakup, and transport by sedimentation, mixing, and advection \citep{pruppacher1978,marley1999b}. Although the modeling of these processes is beyond the scope of this work and will be treated in a future paper, we can speculate on how a sulfur haze subject to microphysics differs from the simple slab model we have used. A major difference would be the vertical profile of the haze. Vertical mixing will loft sulfur particles to higher altitudes, where the lower atmospheric pressure and \seight mixing ratio may result in evaporation of haze particles and a decrease in mean particle size. A haze layer with a vertical gradient in mean particle size would generate a different geometric albedo spectrum than a layer with the same size distribution at all altitudes. 

Given that the sulfur haze would have a source of new particles at or near the altitude where \seight is photochemically produced, there exists the possibility of multi--modal particle size distributions. The situation is similar to that of Venus, where large, $\sim$1 $\mu$m sulfuric acid cloud particles coexist with $\sim$0.1 $\mu$m particles of photochemical origins, possibly composed, at least in part, of elemental sulfur as well \cite[e.g.][]{knollenberg1980,imamura2001,gao2014}. In our case, freshly nucleated sulfur haze particles may coexist with larger, ``aged'' sulfur particles that have grown by the condensation of \seight and other sulfur allotropes, resulting in more complex geometric albedo spectra than those presented here. Setting the particle size distribution width $\sigma$ to 2 in our model is an effort to account for this scenario, but a more detailed study is needed. 

As discussed in ${\S}$\ref{sec:sulfur}, the \seight vapor mixing ratio is largely insensitive to the UV flux and the eddy diffusivity. However, this would no longer be true upon the formation of solid S$_{8}$, as it will lead to \seight rainout. Indeed, while the \seight supersaturation is important in determining the haze mass, another crucial factor is the difference between the haze loss rate due to rainout and the haze production rate, ultimately dependent on the UV flux. For example, if the UV flux were significantly smaller than the nominal values considered in \citet{zahnle2016} and the particle transport were dominated by sedimentation rather than mixing, then any haze particles produced would be lost relatively quickly due to sedimentation and evaporation, leading to a more diffuse haze and depleted \seight abundances above the pressure levels where \seight can condense, much like condensation clouds. On the other hand, if the UV flux and/or eddy diffusivity were high enough such that ample haze particles are produced and are kept aloft in the atmosphere, then we can expect haze masses similar to those shown here. We will examine these processes in more detail in future work.

In addition to the microphysics of the haze itself, the effect of the haze on the rest of the atmosphere must be considered. The chemical abundances presented in Figure \ref{fig:zahnlechem} do not take condensation into account. Indeed, if \seight condenses then all other sulfur allotropes may condense on the \seight particles as well, as their saturation vapor pressures are all significantly lower \citep{lyons2008,zahnle2016}. This will have the effect of not only changing the sulfur chemistry in the atmosphere, possibly removing all sulfur species above the haze layer, but also introduce/increase contamination of the haze particles by smaller sulfur allotropes such as the metastable S$_{3}$ and S$_{4}$. Spectra of these species in the vapor phase have shown absorption bands at 400 nm (S$_{3}$) and 550 nm (S$_{4}$), which would push the UV absorption edge to longer wavelengths compared to what we have shown. This would make giant exoplanets with sulfur hazes redder still, and may allow WFIRST to detect hints of their existence provided that there is a sufficient degree of contamination \citep{meyer1972,meyer1976}. Finally, as H$_{2}$S is key in forming sulfide clouds in exoplanet atmospheres \citep{morley2012}, including the ZnS clouds in our model, and sulfur haze particles can potentially form condensation nuclei for cloud formation, perturbations to the sulfur chemistry and emergence of sulfur hazes could impact condensation clouds as well. 

The strong absorption of UV photons by sulfur hazes is also likely to affect the rest of the atmosphere. On Venus, absorption of UV by the unknown agent in the mesosphere leads to several K day$^{-1}$ of heating \citep{crisp1986,haus2015,haus2016}. Such heating in a giant exoplanet atmosphere may increase temperatures, affecting the sulfur haze abundance. Increasing the haze temperature also increases the wavelength of sulfur's UV absorption edge, though the change is small over the temperature range of relevance \citep[$\sim$0.23 nm K$^{-1}$;][]{meyer1972}. On the other hand, the high albedo generated by the haze at longer wavelengths could decrease atmospheric temperatures by reflecting away more of the stellar flux. Calculating the equilibrium temperature profile due to haze heating and cooling will require a more sophisticated model. 

The current configuration of WFIRST cannot observe the strong absorption of sulfur hazes shortward of 0.45 $\mu$m, which is its defining feature between 0.3 and 1 $\mu$m. Future telescopes capable of direct imaging \citep{dalcanton2015}, such as the LUVOIR \citep{bolcar2016} or HabEx \citep{mennesson2016} mission concepts, as well as ground-based facilities, could potentially probe down to such short wavelengths. Alternatively, sulfur hazes may be discernible by thermal emission spectroscopy in the infrared, which has been conducted on young, bright giant exoplanets by ground based observational campaigns \citep[e.g.][]{macintosh2015,wagner2016,kenworthy2016} and will be especially powerful in the era of JWST. The refractive index of sulfur is essentially featureless between 1 and 2 $\mu$m, and a gap exists in the imaginary component between 2 and 7 $\mu$m, likely indicating that it was too low to be measured precisely. However, at wavelengths longer than 7 $\mu$m, $k$ increases steadily from 10$^{-6}$ to 10$^{-2}$ with punctuating spikes up to as high as 0.1 \citep{fuller1998}, thereby offering possible features by which a sulfur haze could be characterized. In addition, the haze will become optically thin at wavelengths significantly greater than the particle size, such that important molecular features may be detectable in the infrared. We will investigate the effectiveness of JWST and other future telescopes at characterizing giant exoplanets with sulfur hazes in transmission and thermal emission in a future publication.

\acknowledgments

M. S. Marley acknowledges the support of the WFIRST preparatory science program and the NASA XRP program. P. Gao acknowledges the support of the NASA Postdoctoral Program. In addition, we acknowledge and thank the Kavli Summer Program in Astrophysics and its sponsors for providing the opportunity and funding to conduct this research. We thank the Scientific Organizing Committee (J. Fortney, D. Abbot, C. Goldblatt, R. Murray--Clay, D. Lin, A. Showman, and X. Zhang) for putting together an enlightening series of lectures and activities that bolstered our research efforts both at Kavli and at home. We thank P. Garaud, the program coordinator, and others in the Local Organizing Committee (J. Scarpelli, Z. Kornberg, and S. Nasab) for ensuring that our stay at UC Santa Cruz was an amazing experience, being both conducive to research and fun. We thank the senior participants in the Kavli program, in particular D. Catling, K. Menou, and J. Bean, for giving such illuminating and interesting talks to start off our days. Last but far from the least, we thank the Kavli Fellows, A. Baker, C. Leung, C. Cadiou, D. Powell, G. McDonald, K. Feng, K. Ohno, L. Mayorga, M. Malik, N. Batalha, N. Espinoza, R. Garland, F. Ryan, R. MacDonald, T. Komacek, T. Louden, and Y. Kawashima for creating an incredible atmosphere of camaraderie, friendship, adventure, and cake by the ocean.

\vspace{5mm}

\software{FORTRAN, Python, IDL}

\allauthors

\listofchanges

\end{document}